\documentclass[aip,jmp,preprint]{revtex4-1}
\usepackage{amsmath, amssymb}

\newtheorem{theorem}{Theorem}

\newtheorem{corollary}[theorem]{Corollary}
\newtheorem{lemma}[theorem]{Lemma}

\def\qed{\hfill $\Box$\medskip}
\def\IC{{\mathbb C}}
\def\IR{{\mathbb R}}

\def\rank{{\rm rank}\,}

\def\({\left (}
\def\){\right )}

\def\conv{{\rm conv}\,}
 \newcommand{\C}{\mathbb{C}}
 \newcommand{\F}{\mathbb{F}}

 \newcommand{\R}{\mathbb{R}}

 \newcommand{\U}{\mathbf{U}}
 \renewcommand{\u}{\mathbf{u}}

 \newcommand{\x}{\mathbf{x}}
 
 \newcommand{\y}{\mathbf{y}}

 \newcommand{\cG}{\mathcal{G}}

 \newcommand{\cL}{\mathcal{L}}

 \newcommand{\cP}{\mathcal{P}}
 
 \newcommand{\cS}{\mathcal{S}}

 \newcommand{\rC}{\mathrm{C}}
 
 \newcommand{\rH}{\mathrm{H}}
 \newcommand{\rK}{\mathrm{K}}
 \newcommand{\rS}{\mathrm{S}}
 
 \newcommand{\rM}{\mathrm{M}}
 \newcommand{\rD}{\mathrm{D}}
 
 \newcommand{\lan}{\langle}
 \newcommand{\ran}{\rangle}
 \newcommand{\an}[1]{\lan#1\ran}

 \newcommand{\tr}{\mathop{\mathrm{Tr}}\nolimits}

 \newcommand{\PT}{\mathop{\mathrm{PT}}\nolimits}
 \newcommand{\trans}{^\top}
 \newcommand{\proof}{\noindent \textbf{Proof.\ }}

\begin{document}
\openup 1\jot

\title[The automorphism group of separable states]{The automorphism group of separable states in quantum
information theory}

\author{Shmuel Friedland}
\email{friedlan@uic.edu}
\affiliation{Department of Mathematics, Statistics and Computer Science,
University of Illinois at Chicago, Chicago, Illinois 60607-7045, USA}

\author{Chi-Kwong Li}
\email{ckli@math.wm.edu}
\thanks{Li is an honorary professor of the University of Hong Kong, the Taiyuan University of Technology,
and the Shanghai University.}
\affiliation{Department of Mathematics, College of William \& Mary, Williamsburg, VA 23187-8795, USA}

\author{Yiu-Tung Poon}
\email{ytpoon@iastate.edu}
\affiliation{Department of Mathematics, Iowa State University, Ames, IA 50011, USA}

\author{Nung-Sing Sze}
\email{raymond.sze@inet.polyu.edu.hk}
\affiliation{Department of Applied Mathematics,
The Hong Kong Polytechnic University, Hung Hom, Kowloon, Hong Kong}

\date{\today}

\begin{abstract}  
We show that the linear group of automorphism of Hermitian matrices which
preserves the set of separable states is generated by \emph{natural} automorphisms: change of an orthonormal
basis in each tensor factor, partial transpose in each tensor factor, and interchanging
two tensor factors of the same dimension.  We apply our results to
preservers of the product numerical range.
\end{abstract}



\maketitle

\section{Introduction}

One of the main concepts in quantum information theory is \emph{entanglement}.
An entangled state involves at least two subsystem or more.  We first discuss the two subsystem $\rH_m\bigotimes \rH_n$ case, a.k.a. bipartite case. Here $\rM_n$ is the space of $n\times n$ complex matrices
and $\rH_n\subseteq \rM_n$ is the space of $n\times n$ complex Hermitian matrices.
Denote by $\rD_n\subseteq \rH_n$ the convex set of positive semi-definite matrices of trace one, i.e. density matrices.  Also let
$\cS_{m,n}\subseteq \rD_{mn}\subseteq \rH_{mn}\equiv \rH_m \bigotimes \rH_n$
be the set of bipartite separable states, i.e. $\cS_{m,n}=\conv\{A\otimes B: A\in\rD_m\hbox{ and }
B\in\rD_n\}$.  Clearly, $\cS_{m,n}$ is a compact convex set.  The set of entangled bipartite states
is the complement of separable states in $\rD_{mn}$, i.e. $\rD_{mn}\setminus \cS_{m,n}$.

 Among the best known applications of entanglement are superdense coding, quantum teleportation
 and more recently measurement based quantum computation (for review, see e.g. Refs.~\onlinecite{Hor09,Ple07}).
 This recognition sparked an enormous stream of work in an effort to quantify entanglement in both bipartite and multi-partite settings.  Among the different measures of entanglement, the relative entropy of entanglement (REE)  is of a particular importance. The REE
 is defined by (c.f. Ref~\onlinecite{Ved98}):
 \begin{equation*}
 E_{R}(\rho)=\min_{\sigma'\in\mathcal{S}}S(\rho\|\sigma')=S(\rho\|\sigma)\;,
 \end{equation*}
 where $\cS$ is a the set of multi-partite separable states.
 $E_{R}(\rho)$ is a convex function on $\cS$ and is strictly convex on strictly positive
 definite separable states \cite{FG10}.  Hence, the computation of $E_{R}(\rho)$,
 which is given as the minimum of a convex function, should be in principle easy to compute, i.e. polynomial time algorithm \cite{ZFG10}.  However, $E_{R}$ is hard to compute in general, since the general characterization
 of separable states is \emph{NP-hard} \cite{Gur03}.

 A crucial observation of Peres \cite{P} is that $\cS$ is invariant under the partial transpose.
 For example, on $\rH_{mn}\equiv \rH_m\bigotimes\rH_n$ the partial transpose linear map on the second
 component $\PT_2:\rH_{mn}\to\rH_{mn}$ is induced by $\PT_2(A\otimes B)=A\otimes B\trans$, where $B\trans$
 is the transposed matrix of $B\in\rH_n$.
 Hence, if a density matrix $C\in\rD_{mn}$ represents a separable state then $\PT_2(C)$ is positive semi-definite.  (This condition implies that $\PT_1(C)=\PT_2(C)\trans$ is also positive semi-definite, since the transpose map $C\mapsto C\trans$, preserves the trace and the positivity.)
 It was shown in Ref.~\onlinecite{HHH} that for $m+n\le 5$, $C\in \cS_{m,n}$ if and only if $C$ and $\PT_2(C)$ are density matrices.  Unfortunately, the positivity of the partial transpose does not imply separability
 for $m+n\ge 6$ (c.f. Ref.~\onlinecite{HHH}).

 Denote by $\cG(n_1,\ldots,n_k)$ the group of linear automorphisms of Hermitian matrices $\rH_N\equiv \bigotimes_{i=1}^k \rH_{n_i}$ which leaves invariant
 the set of separable states $\cS$.
 The structure of $\cG(m,n)$ was determined recently in Ref.~\onlinecite{AS10}.
 In this paper we extended the above results to $\cG(n_1,\ldots,n_k)$ for $k\ge 3$.  We show that this group is generated by unitary change of basis in each component, partial transposes in each component, and by permutations of the factors of the same dimension.
 In summary, $\cG(n_1,\ldots,n_k)$ consists only of the natural elements.

 There are related works \cite{HPSSSL, Joh10} which study the linear maps on $\otimes_{i=1}^k \C^{n_i}$ that preserve the product states,
 i.e. indecomposable tensors.  In these papers, the authors show some structural results similar to our results on the group $\cG(n_1,\ldots,n_k)$.

 We now briefly summarize the contents of the paper.
 In Section 2, we give another proof for the structure theorem of $\cG(m,n)$
 obtained in  Ref.~\onlinecite{AS10}, and the proof is further extended to determine the structure of
 $\cG(n_1,\ldots, n_k)$ in Section 3. In Section 4, we apply our results to
 preservers of the product numerical range.

 \section{The bipartite case}
 In what follows we use the basic notion of the dimension of a convex set $\rC$ as a subset of $\R^N$, denoted by $\dim\rC$.  It is the minimum of the dimension of an affine space, i.e. a translation of a subspace of $\R^N$, which contains $\rC$.  For a set $\rS\subseteq\R^n$, denote by $\conv \rS$ the convex set spanned by $\rS$.
 For $k$-linear spaces $\U_1,\ldots,\U_k$ over a given field $\F$, we denote by $\bigotimes_{i=1}^k \U_i$
 the tensor vector space of dimension $\prod_{i=1}^k \dim \U_i$.  Suppose $\rS_i$ is a proper subset of $\U_i$ for $i=1,\ldots,k$.  Then
 $$\otimes_{i=1}^k \rS_i=\left\{\otimes_{i=1}^k\u_i: \u_i\in\rS_i, i=1,\ldots,k \right\}.$$

\pagebreak
Denote by $I_m\in\rH_m$ the identity matrix.
Let $\rH_{m}^{+}$ and $\rH_{m}^{(1)}$ denote the set of positive semi-definite matrices
and Hermitian matrices of trace one, respectively.
So $\rH_{m}^{(1)}$ is a hyperplane
in $\rH_m$ with $\dim \rH_{m}^{(1)}=m^2-1$ and $\rD_{m}=\rH_{m}^{+}\cap\rH_{m}^{(1)}$.
Denote by $\cP_m\subseteq \rD_{m}$ the compact set of all Hermitian rank one matrices of trace one,
i.e., the set of pure states.
Then $\cP_m\otimes \cP_n$ is the set of separable pure states in $\rD_{mn}$.
Observe that
$\rK(\cS_{m,n})=\conv(\rH_{m}^+\otimes\rH_{n}^+)
\subseteq \rH_{mn}^+$ is the cone of positive semi-definite
matrices generated by separable states.
The following result is well known and we present the proof for completeness.

\begin{lemma}\label{idintpt}
The set of separable states $\cS_{m,n}$ is a convex set, whose extreme points is
$\cP_m\otimes\cP_n$. Furthermore,
$\dim \cS_{m,n}=(mn)^2-1$ and $\frac{1}{mn}I_{mn}$ is an interior point of $\cS_{m,n}$.
\end{lemma}
\proof
Clearly, since the set of the extreme points of $\rD_m$ is $\cP_m$, it follows that
$\cS_{m,n}=\conv (\cP_m\otimes\cP_n)$.  As $\cP_m\otimes\cP_n\subseteq \cP_{mn}$, it follows that
$\cP_m\otimes\cP_n$ is the set of the extreme points of $\cS_{m,n}$.  Recall next that $\frac{1}{m}I_m$
is an interior point of $\rD_{m}$.  Hence $\frac{1}{mn}I_{mn}=\left(\frac{1}{m}I_m\right)\otimes \left(\frac{1}{n}I_n\right)$
is an interior point of $\cS_{m,n}$.  Since $\cS_{m,n}\subseteq\rH_{mn}^{(1)}$, it follows that  $\dim \cS_{m,n}=(mn)^2-1$. \qed

 \begin{lemma}\label{invcond}  Let $\Phi:\rD_{mn}\to\rD_{mn}$ be an affine map such that $\Phi(\cS_{m,n})=\cS_{m,n}$.  Then $\Phi$ can be extended uniquely to an invertible linear map $\Psi:\rH_{mn}\to\rH_{mn}$.
 \end{lemma}
 \proof   First extend $\Phi$ to an affine homogeneous map, (of degree one), $\Psi:\rK(\cS_{m,n})\to \rK(\cS_{m,n})$ by letting
 $\Psi(tC)=t\Psi(C)$ for any $t\ge 0$ and $C\in\cS_{m,n}$.  Clearly $\Psi$ is affine and homogeneous.
 Also $\Psi(\rK(\cS_{m,n}))=\rK(\cS_{m,n})$.  Since $\rK(\cS_{m,n})-\rK(\cS_{m,n})=\rH_{mn}$,
 it follows that $\Psi$ extends to a linear map of $\rH_{mn}$ to itself.  Since $I_{mn}$ is an interior point of $\rK(\cS_{m,n})$, it follows that $\dim \rK(\cS_{m,n})=(mn)^2$.  Hence, $\dim \Psi(\rK(\cS_{m,n}))=(mn)^2$.
 We claim that the linear map $\Psi$ is invertible.  Otherwise $\dim\Psi(\rH_{mn})\le (mn)^2-1$,
 which contradicts the fact that $\dim \Psi(\rK(\cS_{m,n}))=(mn)^2$.  \qed

 The proof of Lemma \ref{invcond} implies that in order to characterize affine automorphisms of separable bipartite states it is enough to consider linear automorphisms of $\rH_m$ which preserve $\cS_{m,n}$.
 The main result of this section is.

\pagebreak
 \begin{theorem} \label{main1} Let
 $\Psi: \rH_{mn} \rightarrow \rH_{mn}$ be a linear map.
 The following are equivalent.
 \begin{itemize}
 \item[{\rm (a)}] $\Psi(\cP_m \otimes \cP_n) = \cP_m \otimes \cP_n$.
 \item[{\rm (b)}] $\Psi(\cS_{m,n}) = \cS_{m,n}$.
 \item[{\rm (c)}] There are unitary $U \in \rM_{m}$ and $V \in \rM_{n}$ such that

 \medskip
 {\rm (c.1)} $\Psi(A\otimes B) =  \psi_1(A) \otimes \psi_2(B)$
 for $A \otimes B \in \rH_m\bigotimes \rH_n$, or

 {\rm (c.2)}  $m = n$ and $\Psi(A \otimes B) = \psi_2(B)\otimes \psi_1(A)$
 for $A \otimes B \in \rH_m \bigotimes \rH_n$,

 \medskip
 \noindent
 where
 $\psi_1$ has the form $A \mapsto UAU^*$ or $A \mapsto UA\trans U^*$,
 and $\psi_2$ has the form $B \mapsto VBV^*$ or $B \mapsto VB\trans V^*$.
 \end{itemize}
 \end{theorem}

 To prove Theorem \ref{main1}, we need the following lemma which can be viewed as
 the characterization of linear preservers of pure states.

\begin{lemma} \label{lem2}
Suppose $\psi: \rH_m \rightarrow \rH_n$ is linear and satisfies
$\psi(\cP_m) \subseteq\cP_n$. Then one of the following holds:
\begin{enumerate}\renewcommand{\labelenumi}{\rm (\roman{enumi})}
\item there is $R \in \cP_n$ such that $\psi$ has the form
$A \mapsto (\tr A)R$.
\item  $m \le n$ and there is a $U \in \rM_{m\times n}$ with $UU^*  = I_m$ such that $\psi$ has the form
$$ A\mapsto U^*AU \quad\hbox{ or } \quad A\mapsto U^*A\trans U.$$
\end{enumerate}
\end{lemma}

\proof \rm Define a map $\phi: \rH_{m+n} \to \rH_{m+n}$ given by
$$\phi(B) = \phi\left(\begin{bmatrix} B_1 & B_2 \cr B_2^* & B_3 \end{bmatrix}\right)
= \begin{bmatrix} \psi(B_1) & 0 \cr 0 & 0_m \end{bmatrix}
\quad\hbox{for all } B = \begin{bmatrix} B_1 & B_2 \cr B_2^* & B_3 \end{bmatrix} \in \rH_{m+n}
\hbox{ with } B_1 \in \rH_m.$$
Then $\phi$ is linear.
In particular, $\phi(A\oplus 0_n) = \psi(A) \oplus 0_m$ for all $A\in \rH_m$.
Then $\psi(\cP_m) \subseteq \cP_n$ implies $\rank(\phi(A)) \le  1$ whenever $\rank(A) = 1$.
If $\dim \phi(\rH_{m+n})=1$, then there exist a rank one $Q$ and a linear functional $f$ on $\rH_{m+n}$
such that $\phi(B)= f(B)Q$. Therefore, $Q = R\oplus 0_m$ for some $R \in \cP_n$
and $\psi(A) = g(A) R$ for all $A\in \rH_m$ where $g(A) = f(A\oplus 0_n)$.
Since $\psi(\cP_m) \subseteq\cP_n$, $g(P)=1$ for all $P\in \cP_m$.
For $A\in H_m$, let $A=\sum_{i=1}^m\lambda_iP_i$ be the spectral decomposition of $A$. Then
$g(A)= \sum_{i=1}^m\lambda_if(P_i)=\sum_{i=1}^m\lambda_i=\tr A$.

If $\dim \psi(H_m)>1$, by Corollary 2 in Ref.~\onlinecite{BL},
there exist $\alpha \in \{1,-1\}$ and $S\in M_n$ such that $\phi$ has the form
$B\mapsto \alpha S^*BS$ or $B\mapsto \alpha S^*B\trans S$.
Since $\phi(A\oplus 0_n) = \psi(A) \oplus 0_m$, $\psi$ has the form
$$ A\mapsto \alpha U^*AU \quad\hbox{ or } \quad A\mapsto \alpha U^*A\trans U,$$
where is $U$ the leading $m\times n$ submatrix of $S$,
i.e., $S = \begin{bmatrix} U & * \cr * & *\end{bmatrix}$.
Since $\psi(\cP_m) \subseteq \cP_n$, if $\psi$ has the form
$\psi(A) = \alpha U^*AU$, then
$x^*(\alpha UU^*)x = \tr (\alpha U^*(xx^*)U) = \tr( \psi(xx^*)) = 1$
for all unit vector $x \in \IC^m$. This gives  $\alpha UU^* = I_m$.
Hence, $n \ge m$, $\alpha= 1$ and $UU^* = I_m$ and the result follows.
Proof for the case when $\psi(A)=U^*A\trans U$ is similar.
\qed

\medskip\noindent
\bf Proof of Theorem \ref{main1}. \rm
The equivalence of conditions (a) and (b)
follows from the fact that $\cP_m\otimes \cP_n$ is the set of the extreme points of $\cS_{m,n}$
and that $\Psi$ is linear.
The implication  ``(c) $\Rightarrow$ (a)'' is clear.

Suppose (a) holds. 
We will set $\Psi(A\otimes B) = \phi_1(A,B) \otimes \phi_2(A,B)$,
and show that
$(\phi_1(A,B), \phi_2(A,B)) = (\psi_1(A),\psi_2(B))$ for all $A$ and $B$,
or $m = n$ and
$(\phi_1(A,B), \phi_2(A,B)) = (\psi_2(B), \psi_1(A))$ for all $A$ and $B$,
where $\psi_1$ and $\psi_2$ have some standard form.
Below are the technical arguments.

First, Lemma \ref{invcond} yields that $\Psi$ is bijective.
Without loss of generality, we assume that $m \ge n>1$.
Consider the partial traces $\tr_1:\rH_{mn}\to \rH_n$ and $\tr_2:\rH_{mn}\to \rH_m$ on $\rH_{mn}\equiv \rH_m\bigotimes \rH_n$ defined by
$\tr_1(A\otimes B) = (\tr A)\,B$ and $\tr_2(A\otimes B) = (\tr B)\,A$.  Clearly $\tr_1$ and $\tr_2$ are linear maps.
Define two maps
$\phi_1:(\rH_m, \rH_n) \to \rH_m$
and
$\phi_2:(\rH_m, \rH_n) \to \rH_n$ by
$$\phi_1(A,B) = \tr_2(\Psi(A \otimes B))
\quad\hbox{and}\quad
\phi_2(A,B) = \tr_1(\Psi(A \otimes B)).$$
Notice that
\begin{eqnarray}\label{PQ}
\Psi(P \otimes Q) = \phi_1(P,Q) \otimes \phi_2(P,Q)
\quad\hbox{for all}\quad P \in \cP_m \hbox{ and }  Q \in \cP_n.
\end{eqnarray}
Fixed a $Q \in \cP_n$, then the maps $\phi_1(\,\cdot\,,Q): \rH_m \to \rH_m$
and $\phi_2(\,\cdot\,,Q): \rH_m \to \rH_n$ are both linear and
$\phi_1(\cP_m,Q) \subseteq \cP_m$
while $\phi_2(\cP_m,Q) \subseteq \cP_n$.
Therefore, by Lemma \ref{lem2}, both
$\phi_1(\,\cdot\,,Q)$ and $\phi_2(\,\cdot\,,Q)$
have one of the following forms:
\begin{eqnarray}\label{form}
{\rm (i.a)} \quad A\mapsto U^*AU,\quad
{\rm (i.b)} \quad A\mapsto U^*A\trans U,\quad\hbox{or}\quad
{\rm (ii)} \quad A\mapsto (\tr A)\, R,
\end{eqnarray}
where the unitary $U$ and projection $R$ depend on $Q$.
Furthermore, the map $\phi_2(\,\cdot\,,Q)$ can only be of the form (ii) if $m > n$. For $1\le i, j\le m$, let $E_{ij}\in \rM_m$ have $1$ at the $(i,j)$
entry and $0$ elsewhere.
Let $A = E_{11} - E_{22}$. Define
$F: \cP_{n}  \to \IR$ by
$F(Q) = \left\|\phi_1(A,Q) \right\|$,
where $\|\cdot\|$ is the Frobenius norm.
Notice that
$$F(Q) =\left\|\phi_1(A,Q) \right\| =
\begin{cases}
\sqrt 2 & \hbox{if $\phi_1(\,\cdot\,,Q)$
has the form (i.a) or (i.b),} \cr
0 & \hbox{if $\phi_1(\,\cdot\,,Q)$
has the form (ii).}
\end{cases}$$
Now for two distinct $Q_1,Q_2\in \cP_n$, write $Q_1 = \x\x^*$ and $Q_2 = \y\y^*$ with unit vectors $\x,\y \in \IC^n$. Note that  $\x$ and $\y$ are linearly independent. For any $t \in [0,1]$, define
$$Q(t) = \frac{1}{\|\x+t(\y-\x)\|^2}\,
\left(\x+t(\y-\x) \right)(\x+t(\y-\x))^* \in \cP_n.$$
In particular, $Q(0) = Q_1$ and $Q(1) = Q_2$.
For each $t\in [0,1]$,
as $\phi_1(\,\cdot\,,Q(t))$ has the form (i) (i.e. either (i.a) or (i.b)) or (ii), the continuous map
$t\mapsto F(Q(t))$ is constant. Therefore,
one can conclude that either $\phi_1(\,\cdot\,,Q)$ has the form (i) for all $Q\in \cP_m$ or $\phi_1(\,\cdot\,,Q)$ has the form (ii) for all $Q\in \cP_m$.

Now we claim that one of the following holds.
\begin{enumerate}
\item[\rm (I)]  For all $Q\in \cP_n$, $\phi_1(\,\cdot\,,Q)$ has the form (i) and $\phi_2(\,\cdot\,,Q)$
has the form (ii).
\item[\rm (II)] For all $Q\in \cP_n$, $\phi_1(\,\cdot\,,Q)$ has the form (ii) and $\phi_2(\,\cdot\,,Q)$
has the form (i).
\end{enumerate}

Suppose first that for some $Q\in \cP_n$, both $\phi_1(\,\cdot\,,Q)$ and $\phi_2(\,\cdot\,,Q)$ are of the form (i). Then we must have $m =n$.
Then for $r = 1,2$, there is unitary matrix $U_r$ such that
$\phi_r(\,\cdot\,,Q)$ has the form
$A\mapsto U_r^*AU_r$ or $A\mapsto U_r^*A\trans U_r$.
Since $m =n \ge 2$,
the right-hand side of (\ref{PQ}) is a quadratic function in $P\in \cP_m$ while
the left-hand side is linear in $P\in\cP_m$, which is impossible.
To be more precise, let
\begin{equation}\label{P}
P_1 = E_{11},\
P_2 = E_{22},\
P_3 = \frac{1}{2}(E_{11}+E_{12}+E_{21}+E_{22}),
\hbox{ and }
P_4 = \frac{1}{2}(E_{11}-E_{12}-E_{21}+E_{22}).
\end{equation}
Then
$\Psi(P_j\otimes Q) = U^*(P_j\otimes P_j)U$
for all $1\le j \le 4$,
where $U = U_1 \otimes U_2$. Notice that $P_1+P_2 = P_3+P_4$
and hence $P_1\otimes Q + P_2 \otimes Q
= P_3\otimes Q + P_4 \otimes Q$.
But then
$$
\Psi(P_1\otimes Q + P_2\otimes Q)
= U^*( P_1 \otimes P_1 + P_2 \otimes P_2) U
\ne U^*( P_3 \otimes P_3 + P_4 \otimes P_4) U
= \Psi(P_3\otimes Q + P_4\otimes Q),
$$ 
which is a contradiction.

Now suppose that for some $Q\in \cP_n$,  both $\phi_1(\,\cdot\,,Q)$ and $\phi_2(\,\cdot\,,Q)$ are of the form (ii).
Then $\phi_1(A,Q) = (\tr A)\, R_1$ and $\phi_2(A,Q) = (\tr A)\, R_2$
for some $R_1 \in \cP_m$ and $R_2\in \cP_n$.
Therefore, $\Psi(P \otimes Q) = R_1 \otimes R_2$
for all $P \in \cP_m$. This contradicts
the fact that $\Psi$ is a bijective map.
Therefore, either (I) or (II) holds.
Applying a similar argument on the map
$\phi_2(P,\,\cdot\,)$, one can show that
\begin{enumerate}
\item[\rm (III)]  For all $P\in \cP_m$, $\phi_1(P,\,\cdot\,)$ has the form (ii) and $\phi_2(P,\,\cdot\,)$
has the form (i).
\item[\rm (IV)] For all $P\in \cP_m$, $\phi_1(P,\,\cdot\,)$ has the form (i) and $\phi_2(P,\,\cdot\,)$
has the form (ii).
\end{enumerate}

Fix $P_0\in \cP_m$ and $Q_0\in \cP_n$. Suppose (I) and (IV) hold.  Then for any $P\in \cP_m$
and $Q\in \cP_n$,
$$\phi_2(P,Q) = \phi_2(P_0,Q) = \phi_2(P_0,Q_0).$$
Notice that the former equality is by (I) while the latter equality is by (IV).
Contradiction arrived. Similarly, it is impossible that both (II) and (III) hold.
Hence, we can conclude that either (I) and (III) hold or (II) and (IV) hold.

Now suppose (I) and (III) hold. Then
 $\psi_1(\,\cdot\,)=\phi_1(\,\cdot\,,Q_0)$ and  $\psi_2(\,\cdot\,)=\phi_2(P_0,\,\cdot\,)$ are both of the form (i.a) or (i.b).
For all $P\in \cP_m$ and $Q\in \cP_n$,   $\phi_1(P,\,\cdot\,)$ and $\phi_2(\,\cdot\,,Q)$   are both of the form (ii). Hence, $\phi_1(P,Q_0)=\phi_1(P,Q)$ and $\phi_2(P,Q)=\phi_2(P_0,Q)$.
Therefore,
$$\Psi(P\otimes Q)
= \phi_1(P,Q) \otimes \phi_2(P,Q)
= \phi_1(P,Q_0) \otimes \phi_2(P_0,Q)
= \psi_1(P) \otimes \psi_2(Q)
.$$
Then by linearity of $\Psi$ and the fact that $\cP_m \otimes \cP_n$
spans $\rH_{mn}$, the result follows.
Finally, if (II) and (IV) hold, we may replace $\Psi$ by the linear map $A\otimes B \to \Psi(B\otimes A)$ and apply the above argument.
\qed

\section{Extension to multi-partite systems}

One can extend Theorem \ref{main1} to tensor product of
more than two factors as follows:

\begin{theorem} \label{main3}
Suppose $n_1 \ge \cdots \ge n_k\ge 2$ are positive integers with $k > 1$ and
$N=\prod_{i=1}^k n_i$.
Assume that
$\Psi: \rH_N  \rightarrow \rH_N(\equiv \bigotimes_{i=1}^k \rH_{n_i})$ is a linear map.
The following are equivalent.
\begin{itemize}
\item[{\rm (a)}] $\Psi \left( \otimes_{i=1}^k \cP_{n_i} \right) = \otimes_{i=1}^k \cP_{n_i} $.
\item[{\rm (b)}] $\Psi \left(\conv( \otimes_{i=1}^k \cP_{n_i})\right)
= \conv \left(\otimes_{i=1}^k \cP_{n_i} \right)$.
\item[{\rm (c)}] There is a permutation $\pi$ on $\{1,\dots, k\}$
and linear maps $\psi_{i}$ on $\rH_{n_i}$ for $i=1,\dots k$  such that
$$\Psi \left( \otimes_{i=1}^k A_{i} \right) =
\otimes_{i=1}^k \psi_{i}\left(A_{\pi(i)} \right)
\quad \hbox{ for } \ \otimes_{i=1}^k A_k \in \otimes_{i=1}^k \rH_{n_i},$$
where
$\psi_{i}$ has the form $X \mapsto U_{i}XU_{i}^*$ or $X \mapsto U_{i}X\trans \emph{}U_{i}^*$,
for some unitary $U_{i}\in \rM_{n_i}$ and $n_{\pi(i)}=n_i$ for $i=1,\ldots,k$.
\end{itemize}
\end{theorem}

\proof \rm The implications (c) $\Rightarrow$ (a) $\Leftrightarrow$ (b) are clear.
Assume that (a) holds.  A straightforward generalization of Lemma \ref{invcond} yields that $\Psi$ is bijective.
For $1\le r_1 < \cdots< r_p \le k$, define the following linear map
$$\tr^{r_1,\dots,r_p}:
\bigotimes_{i=1}^k \rH_{n_i}\, \to\ \bigotimes_{j=1}^p \rH_{n_{r_j}}
\quad \otimes_{i=1}^k A_i\mapsto
\left( \prod_{i \ne r_1,\dots,r_p} \tr A_i \right) \otimes_{j=1}^p A_{r_j}.$$
In particular, the linear map $\tr^r:\rH_N\to\rH_{n_r}$ is given by
$\tr^{r}  \left( \otimes_{i=1}^k A_{i} \right) =\(\prod_{i\ne r} \tr(A_{i})\) A_{r}$.
For $r = 1,\dots,k$, define maps $\phi_r:(\rH_{n_1},\dots,\rH_{n_k}) \to \rH_{n_r}$ by
$$\phi_r(A_1,\dots,A_k) = \tr^r\left(\Psi\left( \otimes_{i=1}^k A_i \right)\right)
\quad\hbox{for all}\quad
(A_1,\dots,A_k)  \in \left(\rH_{n_1},\dots,\rH_{n_k}\right).$$
Notice that
$$\Psi \left( \otimes_{i=1}^k P_i \right)
= \otimes_{r=1}^k\phi_r(P_1,\dots,P_k)
\quad\hbox{for all}\quad
(P_1,\dots,P_k) \in \left(\cP_{n_1},\dots,\cP_{n_k}\right).$$
Given arbitrary $Q_{i}\in \cP_{n_i}$ for $i = 2, \dots, k$,
the map $\phi_r(\,\cdot\,,Q_2,\dots,Q_k)$
maps $\cP_{n_1}$ into $\cP_{n_r}$.
By Lemma \ref{lem2}, the map must have the form (i) or (ii) in (\ref{form}).
We claim the following.

\medskip\noindent\bf Claim \rm
All but one of the maps
$\phi_r(\,\cdot\,,Q_2,\dots,Q_k)$, $r = 1,\dots,k$,
have the form (ii) for all $Q_i \in \cP_{n_i}$
and the exceptional map has and the form (i)
for all $Q_i \in \cP_{n_i}$.

\medskip
Let $A_1 = E_{11} - E_{22}\in \rH_{n_1}$. Define
$F_r:\left(\cP_{n_2},\dots,\cP_{n_k}\right) \to \IR$ by
$$F_r(Q_2,\dots,Q_k) = \left\|\phi_r(A_1,Q_2,\dots,Q_k) \right\|.$$
Similar to the argument in the proof of Theorem \ref{main1},
$F_r$ is a  constant function. Thus, either

\medskip\hskip .3in
$\phi_r(\,\cdot\,,Q_2,\dots,Q_k)$ always have the form (i) for all $Q_i \in \cP_{n_i}$,
or

\medskip\hskip.3in
$\phi_r(\,\cdot\,,Q_2,\dots,Q_k)$ always have the form (ii)
for all $Q_i \in \cP_{n_i}$.


\medskip
Next, since $\Psi$ is a bijection, it is impossible to have all $\phi_r(\,\cdot\,,Q_2,\dots,Q_k)$ being constant maps.
Assume that the maps $\phi_s(\,\cdot\,,Q_2,\dots,Q_k)$
and $\phi_t(\,\cdot\,,Q_2,\dots,Q_k)$, with $s\ne t$,
have the form (i) and the rest have the form (ii).
In this case, $n_s =n_t = n_1$.
Consider the linear map $L: \rH_{n_1} \to \rH_{n_s} \bigotimes \rH_{n_t}$ defined by
$L(A) = \tr^{s,t} \left(\Psi \left(A\otimes \left(\otimes_{i=2}^k Q_i\right)\right)\right)$.
Then
$$L(P) = \phi_{s}(P,Q_2,\dots,Q_k) \otimes \phi_{t}(P,Q_2,\dots,Q_k)
\quad\hbox{for all}\quad P\in \cP_{n_1}.$$
Recall that $\phi_{s}(P,Q_2,\dots,Q_k)$ and $\phi_{t}(P,Q_2,\dots,Q_k)$
are of the form (i). Following the same argument as in the proof of Theorem \ref{main1}, one sees that
$P_1+P_2 = P_3 + P_4$ while $L(P_1)+L(P_2) \ne L(P_3)+L(P_4)$,
where $P_1,P_2,P_3$, and $P_4$ are defined in (\ref{P}).
This contradicts that $L$ is a linear map. Thus, the claim holds.

For $p = 2,\dots,k$,
applying the same argument on the map
$\phi_r(Q_1,\dots,Q_{p-1},\,\cdot\,,Q_{p+1},\dots,Q_k)$,
one can show that all but one of the map
$\phi_r(Q_1,\dots,Q_{p-1},\,\cdot\,,Q_{p+1},\dots,Q_k)$
have the form (ii) for all $Q_i \in \cP_{n_i}$
and the exceptional map has and the form (i)
for all $Q_i \in \cP_{n_i}$.
\linebreak 
Furthermore, there is a permutation $(\pi(1),\dots,\pi(k))$
of $(1,\dots,k)$ such that
\linebreak 
$\phi_{\pi(p)}(Q_1,\dots,Q_{p-1},\,\cdot\,,Q_{p+1},\dots,Q_k)$
has the form (i) for all $Q_i\in\cP_{n_i}$.
Otherwise, there is $r$ such that
$\phi_{r}(Q_1,\dots,Q_{p-1},\,\cdot\,,Q_{p+1},\dots,Q_k)$
has the form (ii) for all $p$ and for all $Q_i\in\cP_{n_i}$,
which contradicts that $\Psi$ is a bijection.

Notice also that $n_p \le n_{\pi(p)}$ for all $p=1,\dots,k$. This is possible
only when $n_p = n_{\pi(p)}$ for all $p$. Now replacing $\Psi$ by the map of the form
$\otimes_{i=1}^k Q_i \mapsto \Psi\left(\otimes_{i=1}^k Q_{\pi^{-1}(i)} \right)$,
we may assume that $\pi(p) = p$.
Then
$\phi_{p}(Q_1,\dots,Q_{p-1},\,\cdot\,,Q_{p+1},\dots,Q_k)$
has the form (i) for all $Q_i\in\cP_{n_i}$,
and for any $r \ne p$,
$\phi_{r}(Q_1,\dots,Q_{p-1},\,\cdot\,,Q_{p+1},\dots,Q_k)$
has the form (ii) for all $Q_i\in\cP_{n_i}$.
Now fix some $Q_i \in \cP_{n_i}$. Then for any
$P_i \in \cP_{n_i}$,
$$\Psi\left(\otimes_{i=1}^k P_i\right)
= \otimes_{i=1}^k \phi_i(P_1,\dots,P_k)
= \otimes_{i=1}^k \phi_i(Q_1,\dots,Q_{i-1},P_i,Q_{i+1},\dots,Q_k)
= \otimes_{i=1}^k \phi_i(P_i),$$
where $\phi_i(\,\cdot\,) = \phi_i(Q_1,\dots,Q_{i-1},\,\cdot\,,Q_{i+1},\dots,Q_k)$
has the form (i).  By the linearity of $\Psi$, the result follows.
\qed

Next, we show that one cannot replace condition (b) in Theorem \ref{main3} by the weaker
condition that $\Psi$ preserves the separable states  $\cS=\conv( \otimes_{i=1}^k \cP_{n_i})$,
i.e., $\Psi(\cS) \subseteq \cS$.
In fact, we will see that the convex set $\cL$ of separable states preserving linear maps
has dimension $N^4-N^2$, which is the dimension of the convex set of density matrices preserving
linear maps on $\rH_N$.

 \begin{lemma}\label{linseppres}  
Let $\rH_N\equiv \bigotimes_{i=1}^k \rH_{n_i}$. Define the linear map $L_0:\rH_N\to\rH_N$ by
\[L_0(A)=\frac{1}{N}\tr (A) I_N\]
and let $L_1:\rH_N\to\rH_N$ be any linear operator
satisfying
\[\tr (L_1(A))=0 \quad \textrm{for all}\quad A\in \rH_N.\]
 Then there exists $\tau=\tau(L_1)>0$ such that $(L_0+tL_1)(\cS)\subseteq \cS$ for each $t\in (-\tau(L_1),\tau(L_1))$.
 Furthermore $\det (L_0+tL_1)=t^{N^2-1}f(L_1)$, where $f(L_1)$ is a minor of order $N^2-1$ of the representation matrix of
 $L_1$ in a basis of $\rH_N$ which contains $I_N$.
 In particular, if $f(L_1)\ne 0$ then for each $t\in  (-\tau(L_1),\tau(L_1))\setminus\{0\}$ the linear operator $L_0+tL_1$ is invertible.
 \end{lemma}
 \proof Clearly, for each $t\in\R$ the operator $L(t)=L_0+tL_1$ is trace preserving.  Hence it maps the hyperplane $\tr(A)=1$ to itself.
 Note that $L(0)(\cS)=\frac{1}{N}I_N$.  The generalized version of Lemma \ref{idintpt} yields that $\dim \cS=N^2-1$ and $\frac{1}{N}I_N$ is an interior point of $\cS$.
 The continuity argument yields that there exists $\tau=\tau(L_1)$ such that $(L_0+tL_1)(\cS)$ lies in the interior of $\cS$ for $|t|<\tau(L_1)$.

 Let $L_1\trans$ be the adjoint operator of $L_1$ with respect to the standard inner product $\an{A,B}=\tr AB$
 on $\rH_N$.  The assumption that $\tr(L_1(A))=0$ for all $A$ is equivalent to the assumption that $L_1\trans (I_N)=0$.  Note that $L_0\trans=L_0$, $\rank L_0=1$ and $L_0(I_N)=I_N$.
 Choose a basis in $\rH_N$ where $I_N$ is one of the elements of this basis.  Then $L_0=E_{ii}$, for some $i\in\{1,\ldots,N^2\}$ and $L_1$ has a zero row $i$.
 Clearly $\det L(t)=t f(L_1)$ where $f(L_1)$ is corresponding minor of $L_1$.  The last claim of the lemma is obvious.
 \qed

 \begin{corollary}\label{dimseppres}
 Let $\cL$ be the set of all linear transformations $L:\rH_N\to\rH_N$ satisfying $L(\cS)\subseteq \cS$.
 Then $\cL$ is a convex compact set of dimension $N^4-N^2$.  Furthermore the subset $\cL_0\subseteq \cL$ of invertible transformations
 is an open dense set in $\cL$.  Hence $\dim\cL_0=N^4-N^2$.
 \end{corollary}
 \proof
 Since any $L\in \cL$ is trace preserving it follows that $L\trans (I_N)=I_N$.  Let $\cL_1$ be the affine set of all linear transformations
 of $\rH_N$ to itself satisfying
 $L\trans (I_N)=I_N$.  Then $\cL_1$ is a translation of a linear subspace of dimension $N^4-N^2$.
 Hence $\dim\cL\le N^4-N^2$.  Lemma \ref{linseppres} yields  that $\dim \cL=\dim\cL_0=N^4-N^2$.  \qed

 \section{The product numerical range}
 In Ref.~\onlinecite{ZB} the authors introduced the concept of (tensor) product numerical range of
 $T\in \rM_{mn}$ defined by
 \[W^{\otimes}(T)=\{\tr(TX): X\in\cP_{m}\otimes \cP_n\}.\]
 This is also known as the decomposable numerical range associated with the
 tensor product of an operator; see Ref.~\onlinecite{Li} and its
 references.
 It was shown in Refs.~\onlinecite{ZB} and \onlinecite{Ket} that the product numerical range
  is a useful concept in studying various problems in quantum information theory.
 To avoid the nontrivial case we let $m,n\ge 2$.

 Observe that $\rH_m$ is real subspace of $\rM_m$ and $\rM_m=\rH_m\oplus \sqrt{-1}\rH_m$.
 Hence, any real linear automorphism of $\rH_m$ lifts to a complex linear automorphism of $\rM_m$.
 Recall that $\rM_m$ is endowed with the standard inner product $\an{X,Y}=\tr XY^*$.
 Assume that $\Phi:\rM_m\to \rM_m$ is a linear map.
 Then $\Psi^*:\rM_m\to\rM_m$ is the dual linear map given by the equality
 $\an{\Psi(X),Y}=\an{X,\Psi(Y)}$ for all $X,Y\in\rM_m$.
 Theorem \ref{main1} yields.

 \begin{theorem} \label{main2} Let $m,n\ge 2$
 and $\Psi: \rM_{mn}\rightarrow \rM_{mn}$ be a linear map.
 The following are equivalent.
 \begin{itemize}

 \item[{\rm (a)}] $W^{\otimes}(\Psi^*(T))=W^{\otimes}(T)$ for all $T \in \rM_{mn}$.

 \item[{\rm (b)}] $\conv\{W^{\otimes}(\Psi^*(T))\}=\conv\{W^{\otimes}(T)\}$
 for all $T \in \rM_{mn}$.

 \item[{\rm (c)}] $\Psi$ has the form described in Theorem \ref{main1} (c).
 \end{itemize}
 \end{theorem}

 \it Proof. \rm The implications (c) $\Rightarrow$ (a) $\Rightarrow$ (b) are clear.
 Suppose (b) holds.
 Note that
 $$\conv\{W^\otimes(T)\} = \{\tr(TZ): Z \in \cS_{m,n}\}.$$
 Thus the dual map $\Psi^*$ satisfies
 $\Psi^*(\cS_{m,n}) = \cS_{m,n}$ and has the form described in Theorem \ref{main1} (c).
 One readily checks that the dual map of such a map has the same form.
 The result follows. \qed

 In the multi-partite case, we can define the product numerical range of a matrix by
 $$W^\otimes(T) = \left\{\tr(TZ): Z \in \otimes_{i=1}^k \cP_{n_i}\right\},$$
 and deduce the following from Theorem \ref{main3}.

 \begin{theorem} \label{main4}
 Suppose $n_1 \ge \cdots \ge n_k\ge 2$ are positive integers with $k > 1$ and
 $N=\prod_{i=1}^k n_i > 1$.
 Suppose
 $\Psi: \rM_N  \rightarrow \rM_N$ is a linear map.
 The following are equivalent.
 \begin{itemize}

 \item[{\rm (a)}] $W^{\otimes}(\Psi^*(T))=W^{\otimes}(T)$ for all $T \in \rM_N$.

 \item[{\rm (b)}] $\conv\{W^{\otimes}(\Psi^*(T))\}=\conv\{W^{\otimes}(T)\}$
 for all $T \in \rM_N$.

 \item[{\rm (c)}] $\Psi$ has the form described in Theorem \ref{main3} (c).
 \end{itemize}
 \end{theorem}

\section*{Acknowledgments}
This research was done while the second author was visiting the George Washington University during his SSRL leave from the College of William \& Mary in the fall of 2010. Research of Li and Poon was partially supported by USA NSF. Research of Li and Sze was partially supported by HK RGC. Li was also supported by the Key Disciplines of Shanghai Municipality Grant S30104.

 \end{document}